\documentclass[copyright,creativecommons,noderivs,noncommercial]{eptcs}
\providecommand{\event}{ACL2 2011} % Name of the event you are submitting to
\usepackage{breakurl}             % Not needed if you use pdflatex only.

\title{Verifying Sierpi\'{n}ski and Riesel Numbers in ACL2}

\author{
John R. Cowles \qquad\qquad Ruben Gamboa
 \institute{Department of Computer Science \\
            University of Wyoming \\
            Laramie, Wyoming, USA}
 \email{cowles@cs.uwyo.edu \qquad\qquad ruben@cs.uwyo.edu}
}

\def\titlerunning{Verifying Numbers in ACL2}

\def\authorrunning{J.R. Cowles \& R. Gamboa}

\begin{document}
\maketitle

\begin{abstract}
A \emph{Sierpi\'{n}ski number}~\cite[page 420]{pickover} and \cite{filaseta}, 
is an odd positive integer, $k$, such that no positive integer in this infinite
list is prime:
\begin{equation}
  k 2^{1} + 1, k  2^{2} + 1, k 2^{3} + 1, \ldots, k 2^{n} + 1, \ldots .
  \label{eq:sierpinski}
\end{equation}

A \emph{Riesel number}~\cite{filaseta} is similar to a Sierpi\'{n}ski number, with 
$-1$ replacing $+1$ in the above infinite list. Such a number is an odd positive 
integer, $k$, so that no positive integer in this infinite list is prime:
\[ k 2^{1} - 1, k 2^{2} - 1, k 2^{3} - 1, \ldots, k 2^{n} - 1, \ldots . \]

A \emph{cover}, for such a $k$, is a finite list of positive integers such 
that each integer, $j$, in the appropriate infinite list, has a factor, $d$,
in the cover, with $1 < d < j$.

Given a $k$ and its cover, ACL2 is used to systematically verify that each 
integer, in the appropriate infinite list, has a smaller factor in the cover.
\end{abstract}

\section{Introduction}
Sierpi\'{n}ski and Riesel numbers are not easy to find. 
To disqualify an odd positive integer as a Sierpi\'{n}ski number
or a Riesel number, one need only locate a prime in the appropriate infinite list.
With four exceptions, $k = 47, 103, 143, 197$, all of the first 100 odd positive
integers, $1 \le k \le 199$, are disqualified as  Sierpi\'{n}ski numbers by
finding at least one prime in the first eight elements of the infinite 
list~\cite{oeis}:
\[ k 2^{1} + 1, k  2^{2} + 1, k 2^{3} + 1, \ldots, k 2^{8} + 1. \]
Both $k = 103$ and $k = 197$ are eliminated by finding a prime in the list no later
than $k 2^{16} + 1$ \cite{oeis}, leaving $47$ and $143$ as the only possible 
Sierpi\'{n}ski numbers less than $200$. It turns out that $143 \cdot 2^{53} + 1$
and $47 \cdot 2^{583} + 1$ are prime~\cite{oeis}, eliminating them. Thus, there
are no Sierpi\'{n}ski numbers in the range $1 \le k \le 199$. The situation is 
similar for Riesel numbers.

In 1960, W.~Sierpi\'{n}ski~\cite{sierpinski} proved, for 
 \[k = 15511380746462593381, \]
every member in the infinite list, given by (\ref{eq:sierpinski}),
%% \[ k 2^{1} + 1, k  2^{2} + 1, k 2^{3} + 1, \ldots, k 2^{n} + 1, \ldots , \]
is divisible by one of the prime factors of the first six Fermat numbers.
For nonnegative integer, $n$, the \emph{Fermat number}, $F_{n}$, is given by 
  \[ F_{n} = 2^{2^n}+1. \]
The first five Fermat numbers are prime and $F_{5}$ is the product of two primes:
\[ F_{0} = 3, F_{1} = 5, F_{2} = 17, F_{3} = 257, F_{4} = 65537, \]
\[ F_{5} = 4294967297 = 641 \cdot 6700417. \] 
Thus $(3\ 5\ 17\ 257\ 641\ 65537\ 6700417)$ is a cover for 
$k = 15511380746462593381$, showing it to be a Sierpi\'{n}ski number.
Sierpi\'{n}ski's original proof is described in \cite[page 374]{sierpinski1}
and \cite{jones}.

In 1962, J.~Selfridge (unpublished) proved that $78557$ is a Sierpi\'{n}ski 
number by showing that
      \[ (3\ 5\ 7\ 13\ 19\ 37\ 73) \]
is a cover~\cite{17orbust}.
Later, in 1967, Selfridge and Sierpi\'{n}ski conjectured that
$78557$ is the smallest Sierpinski number~\cite{17orbust}.
The distributed computing project Seventeen or Bust~\cite{17orbust} is devoted
to proving this conjecture, disqualifying every $k < 78557$, by finding an $n$
that makes $k \cdot 2^{n} + 1$ prime. For example, $19249 \cdot 2^{13018586} + 1$,
a $3918990$-digit prime, eliminated $19249$~\cite{proth}. When this project 
started in 2002, all but $17$ values of $k$ had already been disqualified. 
Currently six values of $k$ remain to be eliminated.

Earlier, in 1956, but less well known than Sierpi\'{n}ski's work, 
H.~Riesel~\cite{riesel} showed $509203$ is a Riesel number with cover
$(3\ 5\ 7\ 13\ 17\ 241)$. It is possible for the same odd positive integer
to be both a Sierpi\'{n}ski number and a Riesel number.
An example~\cite{filaseta} is $k = 143665583045350793098657$.

\section{Covers Into ACL2 Proofs}
Given an odd positive integer, $k$, with a Sierpi\'{n}ski cover, $\mathcal{C}$,
here is the process used to verify that $k$ is a Sierpi\'{n}ski number. There 
is a similar process for verfying Riesel numbers from their covers.
\begin{enumerate}
 \item For each $d$ in $\mathcal{C}$, find positive integer $b_{d}$ and 
       nonnegative integer $c_{d}$ so that for every nonnegative integer $i$,
       $d$ is a factor of $k \cdot 2^{b_{d} \cdot i + c_{d}} + 1$.

       In practice, every $d$ in $\mathcal{C}$ is an odd prime smaller than $k$.
       \begin{enumerate}
        \item Search for positive integer $b$ such that $d$ is a factor of 
              $2^{b} - 1$. Since $d$ is an odd prime, it turns out that
              such a $b$ will always exist\footnote{For the mathematically 
              literate: The well-known Fermat's Little Theorem ensures the 
              claimed existence.}
              among $1, 2, \ldots, d-1$.
              Let $b_{d}$ be the first\footnote{Thus, being mathematically
              precise, $b_{d}$ is just the order of $2$ in the multiplicative 
              group of the integers modulo $d$.}
              such $b$. \label{induction} 
        \item Search for nonnegative integer $c$ such that $d$ is a factor
              of $k \cdot 2^{c} + 1$. If such a $c$ exists, then one exists
              among $0, 1, \ldots, b_{d}-1$. Let $c_{d}$ be the 
              first\footnote{If $d$ does not divide $k$, then
              $2^{c_{d}} \equiv -(1/k)\pmod d$, so $c_{d}$ is the \emph{discrete
              logarithm}, base $2$, of $-(1/k)$ in the integers modulo $d$.}
              such $c$, if it exists. \label{base}
        \item Assuming $c_{d}$ exists, use induction on $i$, to prove that
              for every nonnegative integer $i$, $d$ is a factor of
              $k \cdot 2^{b_{d} \cdot i + c_{d}} + 1$. \label{result}

              The base case, when $i = 0$, follows from \ref{base} above.

              The induction step, going from $i = j$ to $i = j+1$, follows
              from \ref{induction} above:
%%                \begin{eqnarray}
%%                 \lefteqn{k 2^{b_{d}(j+1) + c_{d}} + 1 =}  \nonumber \\
%%                   & & [k 2^{b_{d}j + c_{d}} \cdot (2^{b_{d}} - 1)]
%%                      + [k 2^{b_{d}j + c_{d}} + 1] \label{eq:sum}
%%                \end{eqnarray}
\begin{equation}
 k 2^{b_{d}(j+1) + c_{d}} + 1 = [k 2^{b_{d}j + c_{d}} \cdot (2^{b_{d}} - 1)]
                              + [k 2^{b_{d}j + c_{d}} + 1] \label{eq:sum}
\end{equation}
              By \ref{induction}, $d$ is a factor of the left summand of 
              (\ref{eq:sum}) and $d$ is a factor of the right summand by
              the induction hypothesis.
       \end{enumerate}
 \item For each positive integer $n$, find $d$ in $\mathcal{C}$ and nonnegative
       integer $i$ so that $n = b_{d} \cdot i + c_{d}$. If such $d$ and $i$
       exist, then, by \ref{result}, $d$ is a factor of 
       $k \cdot 2^{b_{d} \cdot i + c_{d}} + 1 = k \cdot 2^{n} + 1$.  

       To ensure that such $d$ and $i$ exist for every positive $n$, only a
       finite number of cases need be considered: Let $\ell$ be the least 
       common multiple of all the $b_{d}$'s found for the $d$'s in $\mathcal{C}$.
       Check for each 
           \[ n \in \{0, 1, 2, \ldots, \ell - 1 \}, \]
       that there always is a $d$ in  $\mathcal{C}$ that satisfies the equation 
       \[ \bmod(n, b_{d}) = c_{d}. \]
\end{enumerate}
This process has not been formally verified in ACL2. For example, we don't
bother to check that every member of $\mathcal{C}$ is an odd prime. Instead,
for each individual $k$ and $\mathcal{C}$, ACL2 events are generated that would
prove $k$ is a Sierpi\'{n}ski number, if all the events succeed. If some of
the events fail, then, as usual when using ACL2, further study of the failure
is required, in the hope of taking corrective action.  
%% The generation of these events is controlled by the macros: DESCRIBE MACROS
%%% Ruben started writing here.  Originally the following sentence had a : after "macros"
The generation of these events is controlled by the macros
\texttt{verify-sierpinski} and \texttt{verify-riesel}.  These macros
take three arguments: the name of a witness function that will find a
factor for a given $k 2^{n} \pm 1$, the number $k$ that is a
Sierpi{\'{n}}ski or Riesel number, and the cover $\mathcal{C}$ for
$k$.  The macros then generate the proof, following the plan outlined
in this section.

For each $d$ in $\mathcal{C}$, $b_{d}$ and $c_{d}$ from 1a and 1b, are computed.
They are needed to define the witness function and to state the theorems
mentioned in 1c, which are then proved.
For example, the proof that 78557 is a Sierpi{\'{n}}ski number defines
this witness function:
\begin{verbatim}
(DEFUN WITNESS (N)
 (IF (INTEGERP N)
     (COND ((EQUAL (MOD N 2) 0) 3)
           ((EQUAL (MOD N 4) 1) 5)
           ((EQUAL (MOD N 3) 1) 7)
           ((EQUAL (MOD N 12) 11) 13)
           ((EQUAL (MOD N 18) 15) 19)
           ((EQUAL (MOD N 36) 27) 37)
           ((EQUAL (MOD N 9) 3) 73))
     0))
\end{verbatim}
The rightmost numbers, in this definition, form the cover, the corresponding
$b_{d}$'s are the leftmost numbers, and the middle numbers are the $c_{d}$'s.
So  $\mathcal{C} = (3\ 5\ 7\ 13\ 19\ 37\ 73)$, $b_{73} = 9$, and $c_{73} = 3$.

The theorem, from 1c, for $d = 73$ is
\begin{verbatim}
(DEFTHM WITNESS-LEMMA-73
 (IMPLIES (AND (INTEGERP N) 
               (>= N 0))
          (DIVIDES 73
                   (+ 1 
                      (* 78557 
                         (EXPT 2 
                               (+ 3 
                                  (* 9 N)))))))
 :HINTS ...)
\end{verbatim}
Four properties are proved about the witness function, establishing
78557 is a Sierpi{\'{n}}ski number:
\begin{verbatim}
(DEFTHM WITNESS-NATP
 (AND (INTEGERP (WITNESS N))
      (<= 0 (WITNESS N)))
 :HINTS ...)
\end{verbatim}

\begin{verbatim}
(DEFTHM WITNESS-GT-1
 (IMPLIES (INTEGERP N) 
          (< 1 (WITNESS N)))
 :HINTS ...)

(DEFTHM WITNESS-LT-SIERPINSKI
 (IMPLIES (AND (INTEGERP N) 
               (<= 0 N))
          (< (WITNESS N)
             (+ 1 (* 78557 (EXPT 2 N))))))

(DEFTHM WITNESS-DIVIDES-SIERPINSKI-SEQUENCE
 (IMPLIES (AND (INTEGERP N) 
               (<= 0 N))
          (DIVIDES (WITNESS N)
                   (+ 1 (* 78557 (EXPT 2 N)))))
 :HINTS ...)
\end{verbatim}
As suggested above in 2, these properties can be proved by showing
every integer is ``covered'' by one of the cases given in the 
\texttt{COND}-expression used in the definition of the witness function.
\begin{verbatim}
(DEFTHM WITNESS-COVER-ALL-CASES
 (IMPLIES (INTEGERP N)
          (OR (EQUAL (MOD N 2) 0)
              (EQUAL (MOD N 4) 1)
              (EQUAL (MOD N 3) 1)
              (EQUAL (MOD N 12) 11)
              (EQUAL (MOD N 18) 15)
              (EQUAL (MOD N 36) 27)
              (EQUAL (MOD N 9) 3)))
 :RULE-CLASSES NIL
 :HINTS ...)
\end{verbatim}
To prove this, we first demonstrate that these cases are exhaustive when 
$n$ is replaced by $\bmod(n,36)$ (where $36$ is the least common multiple 
of all the moduli above). This can be checked, essentially, by computation.
\begin{verbatim}
(DEFTHM WITNESS-COVER-ALL-CASES-MOD-36
 (IMPLIES (INTEGERP N)
          (OR (EQUAL (MOD (MOD N 36) 2) 0)
              (EQUAL (MOD (MOD N 36) 4) 1)
              (EQUAL (MOD (MOD N 36) 3) 1)
              (EQUAL (MOD (MOD N 36) 12) 11)
              (EQUAL (MOD (MOD N 36) 18) 15)
              (EQUAL (MOD (MOD N 36) 36) 27)
              (EQUAL (MOD (MOD N 36) 9) 3)))
 :RULE-CLASSES NIL
 :HINTS ...)
\end{verbatim}
The actual modular equivalences
that need to be proved depend on both the number 78557 and its cover.
Although the theorem that is being proved is obviously true, there
does not appear to be a way to prove it once and for all in ACL2, not
even using \texttt{encapsulate}.  Instead, a pair of theorems very
much like the ones we have described needs to be proved from scratch
for each different Sierpi{\'{n}}ski or Riesel number.  As experienced
ACL2 users, we are concerned that ACL2 will simply fail to prove this
theorem for some combination of numbers and their covers.  However, we
have used these macros to generate the proof for each of the
Sierpi{\'{n}}ski and Riesel numbers \emph{with covers} listed in the
appendix, and all of the proofs have gone through automatically.
%%Added by John Aug 5 2011:
Note that the appendix essentially\footnote{Given a Sierpi{\'{n}}ski or
Riesel number $k$ and its cover $\mathcal{C}$, infinitely many other examples
can be constructed: Let $P$ be the product of the numbers in $\mathcal{C}$ and 
let $i$ be a positive integer. Then $k + 2 \cdot i \cdot P$ is also a 
Sierpi{\'{n}}ski or Riesel number with the same cover.}
contains all the Sierpi{\'{n}}ski and Riesel numbers known to us.

%%% Ruben stopped writing here

\section{Numbers Without Covers}
There are odd positive integers, shown to be Sierpi\'{n}ski (or Riesel) numbers,
that have no known covers. ACL2 proofs have been constructed for these numbers.

For example~\cite{filaseta}, $k =  4008735125781478102999926000625$ is a 
Sierpi\'{n}ski number, but no (complete) cover is known. For all
positive integer, $n$, if $\bmod( n, 4) \ne 2$, then $k \cdot 2^{n} + 1$ has
a factor among the members of $(3\ 17\ 97\ 241\ 257\ 673)$.
To show $k$ is a Sierpi\'{n}ski number, a factor of $k \cdot 2^{n} + 1$
must be found for all positive integer, $n$, such that $\bmod( n, 4) = 2$. 
Such a factor is constructed using these facts:
\begin{itemize}
 \item $k = 44745755^{4}$ is a fourth power
 \item $4x^{4} + 1 = (2x^{2} + 2x + 1) \cdot (2x^{2} - 2x + 1)$ 
\end{itemize}
Let $i = 44745755$, so $k = i^{4}$. Then
%%  \begin{eqnarray}
%%   k \cdot 2^{4n+2} + 1 & = & 2^{2}(i \cdot 2^{n})^{4} + 1 \nonumber \\
%%                        & = & 4(i \cdot 2^{n})^{4} + 1      \nonumber \\
%%                        & = & (2(i \cdot 2^{n})^{2} + 2(i \cdot 2^{n}) + 1)
%%                                     \label{eq:product} \\  
%%                        &   & \mbox{} \cdot
%%                              (2(i \cdot 2^{n})^{2} - 2(i \cdot 2^{n}) + 1)
%%                              \nonumber
%%  \end{eqnarray}
\begin{eqnarray}
 k \cdot 2^{4n+2} + 1 & = & 2^{2}(i \cdot 2^{n})^{4} + 1 \nonumber \\
                      & = & 4(i \cdot 2^{n})^{4} + 1      \nonumber \\
                      & = & [2(i \cdot 2^{n})^{2} + 2(i \cdot 2^{n}) + 1]
                             \cdot
                            [2(i \cdot 2^{n})^{2} - 2(i \cdot 2^{n}) + 1]
                             \label{eq:product}  
\end{eqnarray}
%% Expression (\ref{eq:product}) algebraically reduces to show
The left factor of (\ref{eq:product}) algebraically reduces to show
%%  \begin{eqnarray*}
%%   \lefteqn{4004365181040050 \cdot 2^{2 \lfloor n / 4 \rfloor}} \\
%%    & & \mbox{} + 89491510 \cdot 2^{\lfloor n / 4 \rfloor} + 1
%%  \end{eqnarray*}
\[ 4004365181040050 \cdot 2^{2 \lfloor n / 4 \rfloor}
   \mbox{} + 89491510 \cdot 2^{\lfloor n / 4 \rfloor} + 1 \]
is a factor of  $k \cdot 2^{n} + 1$, whenever $\bmod( n, 4) = 2$.

A Riesel number, $k$, with no known cover, is given in Appendix A.
In this example, $k = a^{2}$ is a square and
\begin{eqnarray*}
 k \cdot 2^{2n} - 1 & = & a^{2} \cdot 2^{2n} - 1 \\
                    & = & (a2^{n})^{2} - 1 \\
                    & = & (a2^{n} + 1) \cdot (a2^{n} - 1)
\end{eqnarray*}
shows how to factor $k \cdot 2^{m} - 1$ when $m$ is even and positive.
A (partial) cover, listed in Appendix A, gives a constant factor for
each  $k \cdot 2^{m} - 1$, when $m$ is odd and positive.

\section{Conclusions}
Given a Sierpi\'{n}ski or Riesel number, $k$, and its cover, we have described
ACL2 macros that generate events verifying that each integer, in the appropriate
infinite list, has a smaller factor in the cover. 

For the few known Sierpi\'{n}ski or Riesel numbers with no known covers, 
hand-crafted ACL2 proofs have been constructed verifying that each
integer, in the appropriate infinite list, has a smaller factor.

\bibliographystyle{eptcs}
\bibliography{sierpinski}

\appendix
%Appendix A
\section{Sierpi\'{n}ski and Riesel Numbers}
Numbers, $k$, verified with ACL2.

Each $k$ with a cover $\mathcal{C}$ is either mentioned in the References
or claimed at various websites.  Numbers $k$ without known covers are 
from \cite{filaseta}. 
\begin{description}
 \item[Smallest known Sierpi\'{n}ski number] \mbox{ } \\
      $k = 78557 = 17 \cdot 4621$, a product of two primes

      $\mathcal{C} = (3\ 5\ 7\ 13\ 19\ 37\ 73)$
\item[Smallest known prime Sierpi\'{n}ski number] \mbox{ } \\
      $k = 271129$

      $\mathcal{C} = (3\ 5\ 7\ 13\ 17\ 241)$

      \newpage

 \item[More Sierpi\'{n}ski numbers] \mbox{ } \\
      $ \begin{array}{rl} 
   \multicolumn{1}{c}{k} & \ \ \ \ \ \ \mathcal{C} \\ 
   271577  &	(3\ 5\ 7\ 13\ 17\ 241) \\
   322523  &	(3\ 5\ 7\ 13\ 37\ 73\ 109) \\
   327739  &	(3\ 5\ 7\ 13\ 17\ 97\ 257) \\
   482719  &	(3\ 5\ 7\ 13\ 17\ 241) \\
   575041  &	(3\ 5\ 7\ 13\ 17\ 241) \\
   603713  &	(3\ 5\ 7\ 13\ 17\ 241) \\
   903983  &	(3\ 5\ 7\ 13\ 17\ 241) \\
   934909  &	(3\ 5\ 7\ 13\ 19\ 73\ 109) \\
   965431  &	(3\ 5\ 7\ 13\ 17\ 241) \\
   1259779 &	(3\ 5\ 7\ 13\ 19\ 73\ 109) \\
   1290677 &	(3\ 5\ 7\ 13\ 19\ 37\ 109) \\
   1518781 &	(3\ 5\ 7\ 13\ 17\ 241) \\
   1624097 &	(3\ 5\ 7\ 13\ 17\ 241) \\
   1639459 &	(3\ 5\ 7\ 13\ 17\ 241) \\
   1777613 &	(3\ 5\ 7\ 13\ 17\ 19\ 109\ 433) \\
   2131043 & 	(3\ 5\ 7\ 13\ 17\ 241)
     \end{array}$ 
 \item[Smallest Sierpi\'{n}ski number found by Sierpi\'{n}ski] \mbox{ } \\
      $k = 15511380746462593381$

      $\mathcal{C} = (3\ 5\ 17\ 257\ 641\ 65537\ 6700417)$
 \item[Smallest known Riesel number] \mbox{ } \\
      $k = 509203$

      $\mathcal{C} = (3\ 5\ 7\ 13\ 17\ 241)$
 \item[More Riesel numbers] \mbox{ } \\
      $ \begin{array}{rl} 
   \multicolumn{1}{c}{k} & \ \ \ \ \ \ \mathcal{C} \\ 
   762701 & (3\ 5\ 7\ 13\ 17\ 241) \\
   777149 & (3\ 5\ 7\ 13\ 19\ 37\ 73) \\
   790841 & (3\ 5\ 7\ 13\ 19\ 37\ 73) \\
   992077 & (3\ 5\ 7\ 13\ 17\ 241)
     \end{array}$
%%\balancecolumns 
 \item[Numbers both Sierpi\'{n}ski and Riesel] \mbox{ } \\
      $\mathcal{C}_{R}$ indicates the Riesel number cover and
      $\mathcal{C}_{S}$ indicates the Sierpi\'{n}ski number cover.

      \hrulefill \hspace*{\fill}

      $k = 143665583045350793098657$
%%\balancecolumns

      $\mathcal{C}_{R} = (3\ 5\ 13\ 17\ 97\ 241\ 257)$
%%\balancecolumns

      $\mathcal{C}_{S} = (3\ 7\ 11\ 19\ 31\ 37\ 61\ 73\ 109\ 151\ 331\ 1321)$
%%\balancecolumns

      \hrulefill \hspace*{\fill}

      $k = 47867742232066880047611079$

      $\mathcal{C}_{R} = (3\ 7\ 11\ 19\ 31\ 37\ 41\ 61\ 73\ 109\ 151\ 331)$

      $\mathcal{C}_{S} = (3\ 5\ 13\ 17\ 97\ 241\ 257)$

      %%\hrulefill \hspace*{\fill}

      \newpage

      $k = 878503122374924101526292469$

      $\mathcal{C}_{R} = (3\ 7\ 13\ 19\ 37\ 73\ 97\ 109\ 241\ 257)$ 

      $\mathcal{C}_{S} = (3\ 5\ 11\ 17\ 31\ 41\ 61\ 151\ 331\ 61681)$

      \hrulefill \hspace*{\fill}

      $k = 3872639446526560168555701047$

      $\mathcal{C}_{R} = (3\ 7\ 13\ 19\ 37\ 73\ 97\ 109\ 241\ 673)$

      $\mathcal{C}_{S} = (3\ 5\ 11\ 17\ 31\ 41\ 61\ 151\ 331\ 61681)$

      \hrulefill \hspace*{\fill}

      $k = 623506356601958507977841221247$

      $\mathcal{C}_{R} = (3\ 7\ 13\ 19\ 37\ 73\ 97\ 109\ 241\ 673)$

      $\mathcal{C}_{S} = (3\ 5\ 17\ 257\ 641\ 65537\ 6700417)$
 \item[Sierpi\'{n}ski numbers without cover] \mbox{ } \\
      $k =  4008735125781478102999926000625 = 44745755^{4}$

      $(3\ 17\ 97\ 241\ 257\ 673)$ is partial cover for $\bmod( n, 4) \ne 2$.

      $4004365181040050 \cdot 2^{2 \lfloor n / 4 \rfloor}
         + 89491510 \cdot 2^{\lfloor n / 4 \rfloor} + 1 $ \\
      is a factor of  $k \cdot 2^{n} + 1$, whenever $\bmod( n, 4) = 2$.

      \hrulefill  \hspace*{\fill}

      $k = 734110615000775^{4}$

      $(3\ 17\ 257\ 641\ 65537\ 6700417)$ is partial cover for 
      $\bmod( n, 4) \ne 2$.

      $1077836790113632192906501201250 \cdot 2^{2 \lfloor n / 4 \rfloor}
       \mbox{} + 1468221230001550 \cdot 2^{\lfloor n / 4 \rfloor} + 1$ \\
      is a factor of  $k \cdot 2^{n} + 1$, whenever $\bmod( n, 4) = 2$.
 \item[Riesel number without cover] \mbox{ } \\
%%\balancecolumns
      Let 
%%  $a = \mbox{\scriptsize 3896845303873881175159314620808887046066972469809}$
 $a = 3896845303873881175159314620808887046066972469809$ 
     and let $k = a^{2}$.

     The list

     $(7\ 17\ 31\ 41\ 71\ 97\ 113\ 127\ 151\ 241\ 257\ 281\ 337
       \ 641\ 673\ 1321\ 14449\ 29191\ 65537\ 6700417)$ \\
     is partial cover for odd positve $n$.

      $a \cdot 2^{n/2} + 1$ is a factor of $k \cdot 2^{n} - 1$, whenever
      $n$ is positive and even.
\end{description}

\end{document}